\documentclass[aps,twocolumn, amsmath,amssymb, superscriptaddress, reprint]{revtex4-1}

\usepackage[utf8]{inputenc}
\usepackage{lmodern}
\usepackage[version=3]{mhchem}
\usepackage{amsmath,amssymb,amstext,amsfonts}
\usepackage{natbib}
\usepackage{psfrag,graphicx}
\usepackage[]{units}
\usepackage{upgreek}
\usepackage{textcomp} 
\usepackage{color}
\usepackage{float}
\usepackage{subfigure}
\usepackage{array} % Matrizen in mathematischen Formeln
\usepackage{epstopdf}
\usepackage[ngerman, num]{isodate}

\begin{document}
% ------
% Maketitle metadata
\title{3D Printing of Polymer Bonded Rare-Earth Magnets With a Variable Magnetic Compound Density for a Predefined Stray Field}

\author{C. Huber}
\thanks{Correspondence to: \href{mailto:christian.huber@tuwien.ac.at}{christian.huber@tuwien.ac.at}}
\affiliation{Institute of Solid State Physics, Vienna University of Technology, 1040 Vienna, Austria}
\affiliation{Christian Doppler Laboratory for Advanced Magnetic Sensing and Materials, 1040 Vienna, Austria}

\author{C. Abert}
\affiliation{Institute of Solid State Physics, Vienna University of Technology, 1040 Vienna, Austria}
\affiliation{Christian Doppler Laboratory for Advanced Magnetic Sensing and Materials, 1040 Vienna, Austria}

\author{F. Bruckner}
\affiliation{Institute of Solid State Physics, Vienna University of Technology, 1040 Vienna, Austria}
\affiliation{Christian Doppler Laboratory for Advanced Magnetic Sensing and Materials, 1040 Vienna, Austria}

\author{M. Groenefeld}
\affiliation{Magnetfabrik Bonn GmbH, 53119 Bonn, Germany}

\author{S. Schuschnigg}
\affiliation{Department of Polymer Engineering and Science, Montanuniversitaet Leoben, 8700 Leoben, Austria}

\author{I. Teliban}
\affiliation{Magnetfabrik Bonn GmbH, 53119 Bonn, Germany}

\author{C. Vogler}
\affiliation{Institute of Solid State Physics, Vienna University of Technology, 1040 Vienna, Austria}

\author{G. Wautischer}
\affiliation{Institute of Solid State Physics, Vienna University of Technology, 1040 Vienna, Austria}
\affiliation{Christian Doppler Laboratory for Advanced Magnetic Sensing and Materials, 1040 Vienna, Austria}

\author{R. Windl}
\affiliation{Institute of Solid State Physics, Vienna University of Technology, 1040 Vienna, Austria}
\affiliation{Christian Doppler Laboratory for Advanced Magnetic Sensing and Materials, 1040 Vienna, Austria}

\author{D. Suess}
\affiliation{Institute of Solid State Physics, Vienna University of Technology, 1040 Vienna, Austria}
\affiliation{Christian Doppler Laboratory for Advanced Magnetic Sensing and Materials, 1040 Vienna, Austria}

%\date{16.11.2016}
\date{\today}

$ $\newline
%%%%%%%%%%%%%%%%%%%%%%%%

%\thispagestyle{empty}

\begin{abstract}
Additive manufacturing of polymer bonded magnets is a recently developed technique, for single-unit production, and for structures that have been impossible to manufacture previously. Also new possibilities to create a specific stray field around the magnet are triggered. The current work presents a method to 3D print polymer bonded magnets with a variable magnetic compound density distribution. A low-cost, end-user 3D printer with a mixing extruder is used to mix permanent magnetic filaments with pure PA12 filaments. The magnetic filaments are compounded, extruded, and characterized for the printing process. To deduce the quality of the manufactured magnets with a variable compound density, an inverse stray field framework is used. The effectiveness of the printing process and the simulation method is shown. It can also be used to manufacture magnets that produce a predefined stray field in a given region. Examples for sensor applications are presented. This setup and simulation framework allows the design and manufacturing of polymer bonded permanent magnets which are impossible to create with conventional methods. 
\end{abstract}

\maketitle

\section{Introduction}
Additive manufacturing is an affordable, rapid technique to manufacture models, tools, prototypes, or end products. The production is carried out directly from formless (liquids, powders, etc.) or form-neutral (tape, wire) material mostly by means of thermal or chemical processes. No specific tools are required for a specific object with a possibly complex shape. A well established additive manufacturing method is the fused deposition modeling (FDM) technology. FDM, also referred to as 3D printing, is a process that uses wire-shaped thermoplastic filaments. The filament is heated just above its softening point with the aid of a moving heated extruder. Molten thermoplastic is pressed out of the printer head nozzle and builds up the object layer by layer on the already solidified material on the building platform \cite{3d-print}. Since 3D printers are nowadays affordable for end-users, a boom of new possibilities have been triggered \cite{wohlers}. 3D print technology is a fast growing field for single-unit production, and it allows to produce structures that have been difficult or impossible to build before.

NdFeB magnets are mainly divided into sintered and polymer bonded magnets. On the one hand, sintered magnets have the highest maximum energy product $(BH)_{\text{max}}$, on the other hand polymer bonded magnets enable the manufacturing of complex shapes and magnetization structures, but with a lower $(BH)_{\text{max}}$ \cite{recent_devel}. Therefore, they are  widely used wherever product cost is a major consideration over magnetic performance \cite{bonded_mag_future}. Bonded magnets offer a wide application range from sensor to actuator applications \cite{ibb}. 

Polymer bonded magnets are composites with permanent-magnet powder embedded in a polymer binder matrix. Hard magnetic particles, ferrite (e.g. Sr, Ba), and rare-earth materials (e.g. NdFeB) with a volume filler content between 40~--~65\,vol.\%  are inserted. These compounds can be further processed with injection molding or extrusion \cite{diss_drummer}. The NdFeB particles for the compounds are produced by a melt spinning process. To achieve better rheological behavior, spherical particles are preferred, which can be produced by an inert gas atomization process. To reduce assembling costs and reach more flexibility, magnetically isotropic powder is preferred. The high filler content increases the viscosity of the melted compound \cite{models_visc}. To avoid clogging of the nozzle, the matrix polymer should be of a high flowable material, good mechanical properties are an important aspect, too. Polyamide (PA6, PA11, and PA12) have a good combination of these qualities.

Recently it was shown that an end-user 3D printer can be used to print polymer bonded rare-earth magnets with a complex shape \cite{pub_16_1_apl}. A prefabricated magnetic compound (Neofer\,\textregistered$ $ 25/60p) from Magnetfabrik Bonn GmbH has been used. It consists of 90\,wt.\% NdFeB grains in a PA11 matrix. The effectiveness of this printing method is demonstrated by fabrication of a magnet with a complex shape that is known to produce a specific stray field above the printed magnet. Structures with a size of under 0.8\,mm, and a layer height of under 0.1\,mm are possible. Contrary to the well established, affordable, accurate, and high resolution end-user 3D printing technology, big area additive manufacturing (BAAM) of large scale NdFeB magnets is presented in \cite{baam}. The BAAM method operates within the same principle as a conventional 3D printer. An advantage of this method is the possibility to manufacture large scale objects, disadvantages are the high system costs and printing of fine structures is impossible due to the large printer nozzle size.

However, at the moment no other single-unit manufacturing technologies are available for the production of magnets with complex shapes, as well as the opportunity to fabricate objects without material waste, and a minimum amount of source material. This can be an important aspect for the reduction of rare-earth elements in permanent magnets \cite{reduction_ndfeb}.

In this work, a method to produce polymer bonded permanent magnets with a variable magnetic compound density along the printing direction is presented. The magnetic powder filler fraction $\varrho_M$ is proportional to the remanence $B_r$. This can be used to shape the magnetic field without changing the topology of the object. First, the effectiveness of the method is shown. Furthermore an inverse stray field method based on finite elements has been developed, which allows to deduce the compound density and magnetization distribution of the magnet from stray field measurements. This method can be used to evaluate the quality of the printed magnets. Moreover, the inverse method allows us to find an optimal magnetization density distribution for a given target field.

\section{Results}
\subsection{Predefined Magnetic Compound Density}
\begin{figure*}[ht]
	\centering
	\includegraphics[width=1\textwidth]{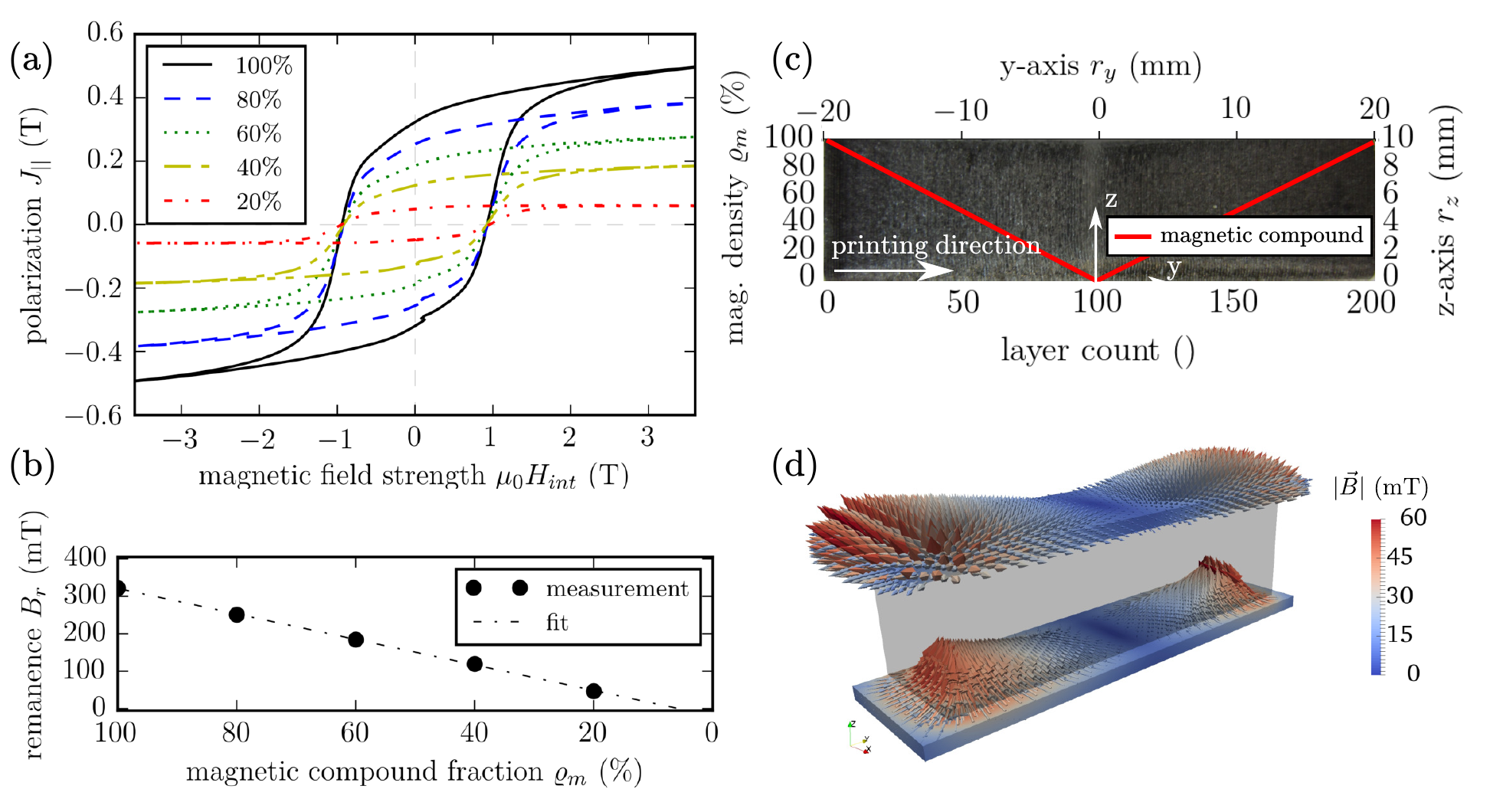}
	\caption{3D print of polymer bonded magnets with a variable magnetic compound density. (a) Hysteresis measurements of the permanent magnetic powder (MQP-S-11-9) inside the PA12 matrix with a variable magnetic compound fraction $\varrho_m$. (b) Linear declining remanence $B_r\sim-\varrho_m$. (c) Picture of the printed cuboid (10$\times$40$\times$10\,mm$^3$ (L$\times$W$\times$H)), and the magnetization distribution along the $y$-axis $r_y$. (d) Volume scan of the produced stray field above and under the printed magnet.}
	\label{fig:triangular_print}
\end{figure*}
A mixing extruder of an end-user 3D printer has the possibility to mix two or more materials during the printing process. In this article the mixing extruder is used to mix magnetic compound material with pure commercial PA12. The magnetic compound consist of 85\,wt.\% NdFeB particles inside a PA12 matrix. Commercial magnetically isotropic powder MQP-S-11-9 with the chemical composition NdPrFeCoTiZrB from Magnequench Corporation is used. The powder is produced by employing an atomization process followed by heat treatment. The particles are of spherical morphology with a diameter of approximately 45$\pm$20\,\textmu m (Supplement Fig.~1). The magnetic compound is extruded into suitable filaments with a diameter of 1.75$\pm$0.1\,mm and a magnetic filler content of 85\,wt.\% and 43\,vol.\%, respectively.

The mixing extruder can continuously change between both materials. The magnetic compound density is a function of the layer number and $y$-axis $r_y$, respectively. To determine the magnetic properties of the prints with different magnetic filler fractions, hysteresis measurements are performed and pictured in Fig.~\ref{fig:triangular_print}(a). Volumetric mass density measurements yield $\varrho=3.2$\,g/cm$^3$ for the maximum magnetic filler fraction of $\varrho_m=100$\,\%. This is 15\,\% lower as the theoretical density of the compound. The compound exhibits a remanence $B_r=314$\,mT and a coercivity $H_{cj}=745$\,kA/m. However, the remanence $B_r$ decreases linearly with the magnetic compound density $\varrho_m$ (Fig.~\ref{fig:triangular_print}(b)). This means that the maximum energy product ($(BH)_{\text{max}}\sim B^2$) is proportional to $\varrho_{m}^2$. To benchmark the variable magnetic compound printing method, a cuboid of size  10$\times$40$\times$10\,mm$^3$ (L$\times$W$\times$H) with an absolute value magnetic density function ($\varrho_m=100\,\%/(W/2)|r_y|\,\%$) is printed (Fig.~\ref{fig:triangular_print}(c)). The sample is magnetized inside an electromagnet with 1.9\,T along the $z$-axis. A volume scan of the produced stray field above and under the magnetized cuboid is pictured in Fig.~\ref{fig:triangular_print}(d) \cite{pub_16_1_apl}. This measurement will be used to reconstruct the magnetization distribution inside the magnet and therefore, deduct the quality of the printed magnet.

\subsection{Inverse Problem}

The forward stray field computation problem is defined by finding the stray field for a given magnetization. Well established finite element method (FEM) algorithms for the stray field calculation of permanent magnets  exists \cite{stray_field_fem}. In contrast to the forward problem, the inverse problem, where for a given magnetic field outside the magnet the magnetization within the magnet is reconstructed, is much harder to solve (Supplementary Fig.~2). The inherently difficulty of this inverse problem is due to the facts that (i) the inverse problem is not unique (ii) the underlying system of equations is ill-conditioned.
%To deduct the quality of the printed structures or to find a magnetization distribution for a specific target stray field a calculation method is required which input is only the geometry and the target stray field in a defined region outside the magnet.
%A forward stray field calculation calculate the stray field from a given magnetization. It exists several well known finite element method (FEM) algorithms for the stray field calculation of permanent magnets \cite{stray_field_fem}. Compared with the forward problem, the inverse stray field computation where the magnetization is the result, is regarding their worse conditions harder to solve  (Supplementary Fig. 2).
Mostly, no unique solution is available for these kind of problems. A method to solve the inverse problem by using an adjoint method exists \cite{fredkin, inverse_flo}. In this article, a pure FEM method based on the FEM library FEniCS \cite{fenics}, and the library dolfin-adjoint \cite{dolfin} for the automatically derivation of the adjoint equation of a given forward problem. Dolfin-adjoint contains a framework to solve partial differential equation (PDE) constraint optimization problems.

The forward problem is a well-posed problem. This means a solution exists and it is unique. As above-mentioned, the inverse problem is ill-posed. To provide an approximated solution of the inverse problem additional informations are necessary. Different methods exists to find reasonable results \cite{inverse}. Here, the Tikhonov regularization is implemented in the inverse stray field computation framework. Solving the following minimization problem results in the unknown magnetization $\vec{M}$ for each finite element of the model in the region $\Omega_m$ (Supplementary Fig.~3):

\begin{align}
\label{eq:min}
 %\min_{\vec{M}}\Arrowvert \vec{h_{sim}} - \vec{h_{exp}}\Arrowvert^2_2 + \underbrace{\alpha\Arrowvert \nabla \vec{M}\Arrowvert^2_2}_{\mathrm{regularization}}
   \min_{\vec{M}} \left(\int_{\Omega_{h}} \Arrowvert \vec{h_{\text{sim}}}-\vec{h_{\text{exp}}}\Arrowvert^2 \mathrm{d} \vec{r} + \underbrace{\alpha\int_{\Omega_{m}}\Arrowvert \nabla \vec{M}\Arrowvert^2 \mathrm{d} \vec{r}}_{\mathrm{regularization}}\right)
\end{align}

where $\vec{h_{\text{sim}}}$ is the stray field calculated by the forward problem in a defined region $\Omega_h$, with the magnetic potential $u$.  $\vec{h_{\text{exp}}}$ is the measured or target stray field in the same region $\Omega_h$. $\alpha\geqslant 0$ is the Tikhonov regularization parameter. In this case $\alpha$ has units m$^2$. 

The main challenge for this regularization is the proper choose of a suitable parameter $\alpha$. If $\alpha$ is too small the solution will be dominated by the contributions from the data errors. If $\alpha$ is too large the solution is a poor approximation of the original problem. A well-known method to find an optimal $\alpha$, is the so-called L-curve method \cite{l-curve}. For this method the solution norm $\Arrowvert \nabla \vec{M}\Arrowvert^2_2$ is plotted over the residual norm $\Arrowvert \vec{h_{\text{sim}}} - \vec{h_{\text{exp}}}\Arrowvert^2_2$ in a log-log scale for varying $\alpha \in [0,\infty)$. The optimal residual parameter $\alpha$ is where the curve has the maximum curvature (corner of the L-curve). This $\alpha$ value gives a good compromise between the change of the residual norm and reduction of the solution norm (Supplementary Fig.~4). To solve the minimization problem in Eq.~\ref{eq:min}, the IPOPT software library for large scale nonlinear optimization systems is used \cite{ipopt}.

\subsection{Reconstructed Magnetization}
\begin{figure*}[ht]
	\centering
	\includegraphics[width=1\textwidth]{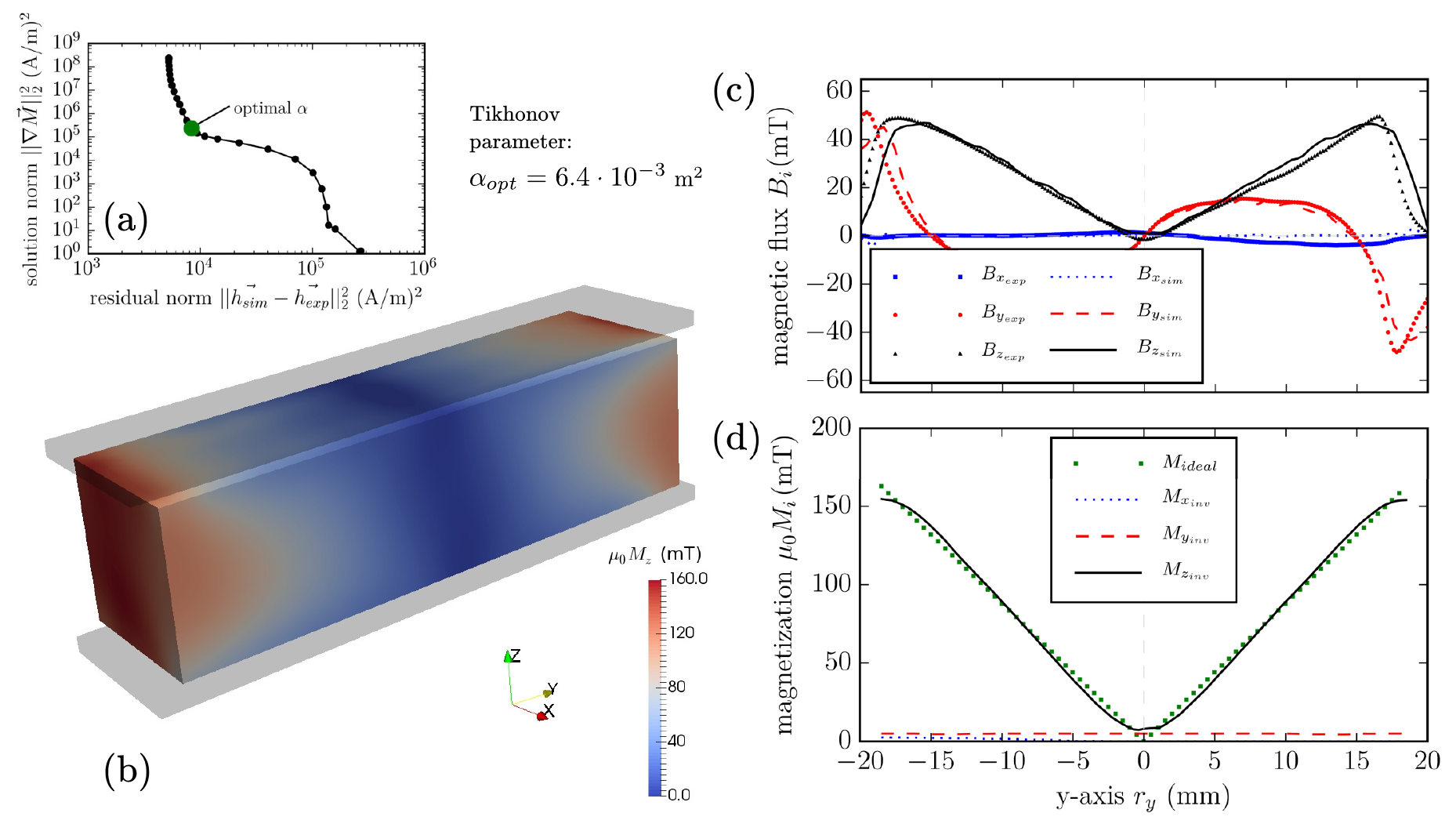}
	\caption{Reconstructed magnetization of a cuboid printed structure. (a) L-curve to find the optimal Tikhonov regularization parameter $\alpha$. (b) Reconstructed magnetization distribution $\mu_0M_z$ of the magnet. (c) Line scan of the stray field 1.5\,mm above the magnet compared with the inverse stray field simulation results. (d) Ideal magnetization in the middle of the magnet along the $y$-axis $r_y$ compared with the reconstructed magnetization distribution.}
	\label{fig:alpha_sweep_opt}
\end{figure*}
To benchmark the inverse stray field framework, and deduce the quality of the 3D printed magnetic cuboid with an absolute value magnetic density distribution along the $y$-axis the printed and magnetized magnet is scanned on both sides in a volume of 40$\times$12$\times$2\,mm$^3$ (L$\times$W$\times$H) with a spatial resolution of 0.2\,mm in the magnetization direction $r_z$ (Fig.~\ref{fig:triangular_print}(d)). The measured $\vec{h_{\text{exp}}}$ is the input for the inverse stray field calculation. The simulation is performed for a range of different Tikhonov regularization parameters $\alpha=10^{x}$\,m$^2$ with $x \in [-9.4,3]$ and a step size of 0.4. Fig.~\ref{fig:alpha_sweep_opt}(a) shows the L-curve with the different $\alpha$ values, and the optimal solution with $\alpha_{\text{opt}}=6.4\cdot10^{-3}$\,m$^2$. Regarding the error of the measurement, the L-curve looks different to the ideal one (Supplementary Fig.~4). However, the criteria for an optimal $\alpha$ is obviously fulfilled. Fig.~\ref{fig:alpha_sweep_opt} (b) illustrates the magnetization distribution $M_z$, which is proportional to the magnetic density distribution inside the magnet. A line scan 1.5\,mm above the magnet compared with the simulation results are shown in Fig.~\ref{fig:alpha_sweep_opt}(c). It points out a good agreement between measurement and results from the inverse stray field calculation. The distribution along the $y$-axis in the middle of the magnet is plotted in Fig.~\ref{fig:alpha_sweep_opt}(d). The reconstructed magnetization $M_{z_{\text{inv}}}$ fits very well with the ideal magnetization distribution of $M_{\text{ideal}}=M_{\text{max}}/(W/2)|r_y|$\,mT or $\varrho_m=100\,\%/(W/2)|r_y|$\,\% for the magnetic density distribution. Where $M_{\text{max}}$ is the maximum magnetization and $W$ is the width of the magnet. The reconstructed components $M_{x_{\text{inv}}}$ and $M_{y_{\text{inv}}}$ are small compared to the z component. This complies with the expectations of the printed permanent magnet. A supplementary animation shows the change of the magnetization distribution and the resulting stray field at different Tikhonov regularization parameters $\alpha$. If $\alpha \rightarrow 0$, the magnetic compound density $\varrho_m$ distribution is unphysical, but the stray field fits with the measurement data. If  $\alpha \rightarrow \infty$,  $M_{z_{\text{inv}}}=1$\,mT for the whole magnetic region and therefore, the stray field above the magnet mismatch with the measurement data.

\subsection{Predefined Stray Field}
\begin{figure*}[ht]
	\centering
	\includegraphics[width=1\textwidth]{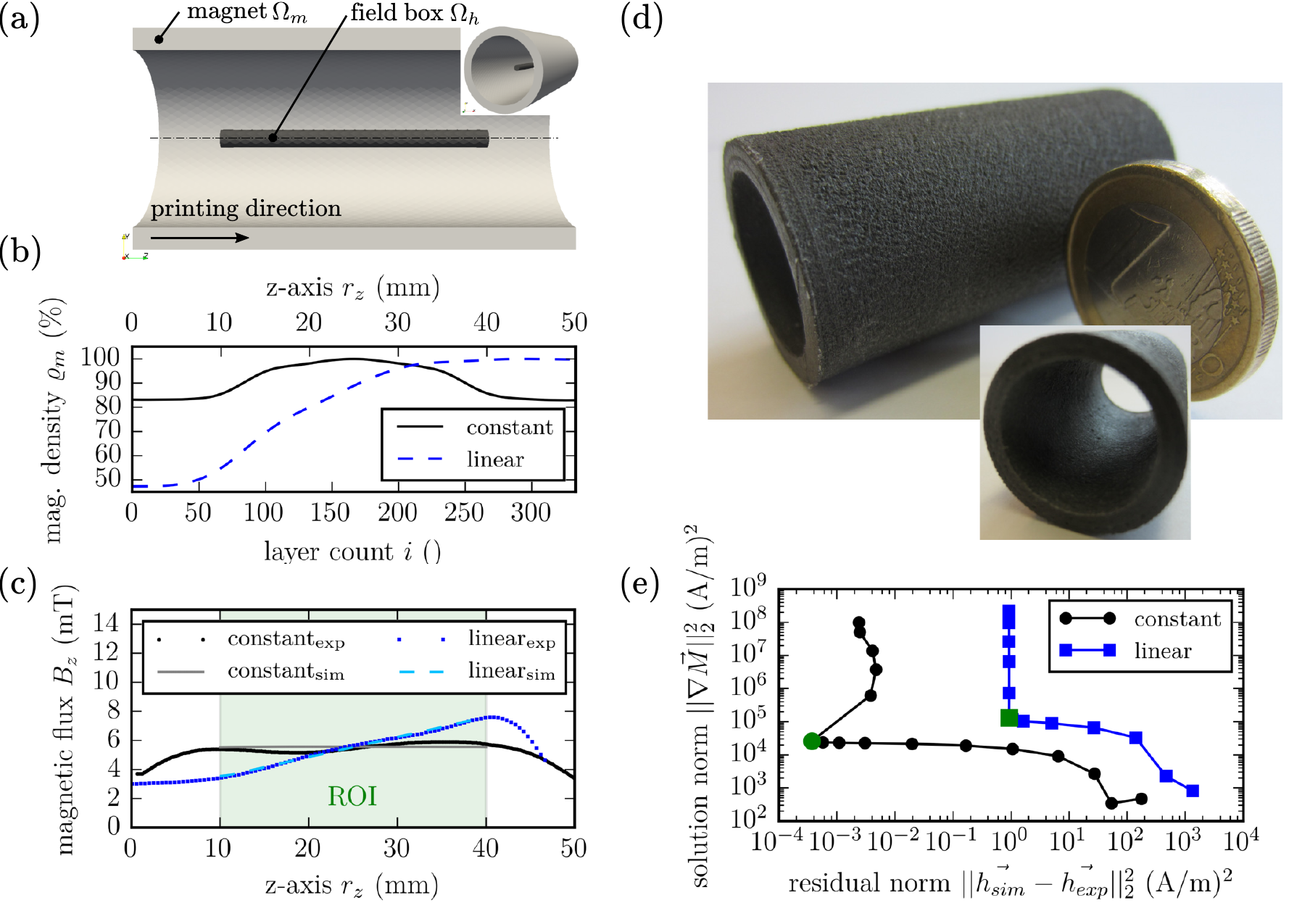}
	\caption{3D prints of magnetic hollow cylinder with a variable magnetic compound density distribution to generate a predefined stray field inside the cylinder. (a) Model of the hollow cylinder magnet with the dimension in mm ($\varnothing$25,~$\varnothing$20,~50 ($d_{\text{outer}},~d_{\text{inner}},~L$)) with a predefined stray field in the field box ($\varnothing$2,~30 ($d,~L$)). (b) Magnetic compound density distribution $\varrho_m$ along the $z$-axis $r_z$ to create a constant and linear stray field in the field box, respectively. (c) Stray field measurements of $B_z$ compared with inverse stray field FEM simulations in the middle of the hollow cylinder for the linear and the constant field generations magnet, respectively. (d) Picture of the hollow cylindrical magnet. (e) L-curve for both designs to find the optimal Tikhonov regularization parameter $\alpha$.}
	\label{fig:density_print}
\end{figure*}
Instead of using the inverse stray field method to investigate already printed magnets, the method can also be used to design magnets with a specific stray field properties. As examples we compute optimal magnetization distribution for a hollow cylinder geometry for different target fields inside the cylinder.  The hollow cylinders have the dimension in mm $\varnothing$25,~$\varnothing$20,~50 ($d_{\text{outer}},~d_{\text{inner}},~L$) with a linear and a constant stray field distribution inside the hollow magnet. Fig.~\ref{fig:density_print}(a) shows the model of the magnet with the magnetic region $\Omega_m$ and the region for the predefined stray field $\Omega_h$. The printing direction is along the $z$-axis. For this reason, the variable $\vec{h_{\text{exp}}}$ in Eq.~\ref{eq:min} does not represent the measurement data but rather the desired stray field in distributions $\Omega_h$. $M_x$ and $M_y$ is fixed to zero, and the maximum of $M_z$ is limited to the used magnetic material.
Otherwise the real printed magnet can not reach the desired magnetization. Two different stray field distributions are tested. The first is a constant magnetic flux density of $B_z=5.5$\,mT along the $z$-axis $r_z \in [10,40]$\,mm, the second one is a linear increasing field of $B_z=2+0.15r_z$\,mT/mm along the $z$-axis $r_z \in [10,40]$\,mm. A constant magnetic field inside a hollow cylinder can be used to calibrate sensors where the sensor position is changing. A linear increasing field can be used to realize a linear positioning system. In this case, a 1D sensor is enough for an accurate position detecting system \cite{linear}. The inverse stray field simulation for both examples are performed for various Tikhonov regularization parameter $\alpha=10^{x}$\,m$^2$ with $x \in [-10,1]$. The L-curve for both simulations are presented in Fig.~\ref{fig:density_print}(e). $\alpha_{\text{opt}}$ is clearly visible and is marked in green ($\alpha_{\text{opt}}=2.5\cdot10^{-7}$\,m$^2$ for both designs). The resulting magnetic density distribution along the $z$-axis is plotted in Fig.~\ref{fig:density_print}(b). Fig.~\ref{fig:density_print}(c) shows the comparison between simulations and measurements in the middle of the hollow cylinders. Inside the field boxes with the dimensions $\varnothing$2,~30\,mm ($d,~L$) a good conformity between printed and simulated magnets is given.
\begin{figure}[htbp]
	\centering
	\includegraphics[width=0.8\linewidth]{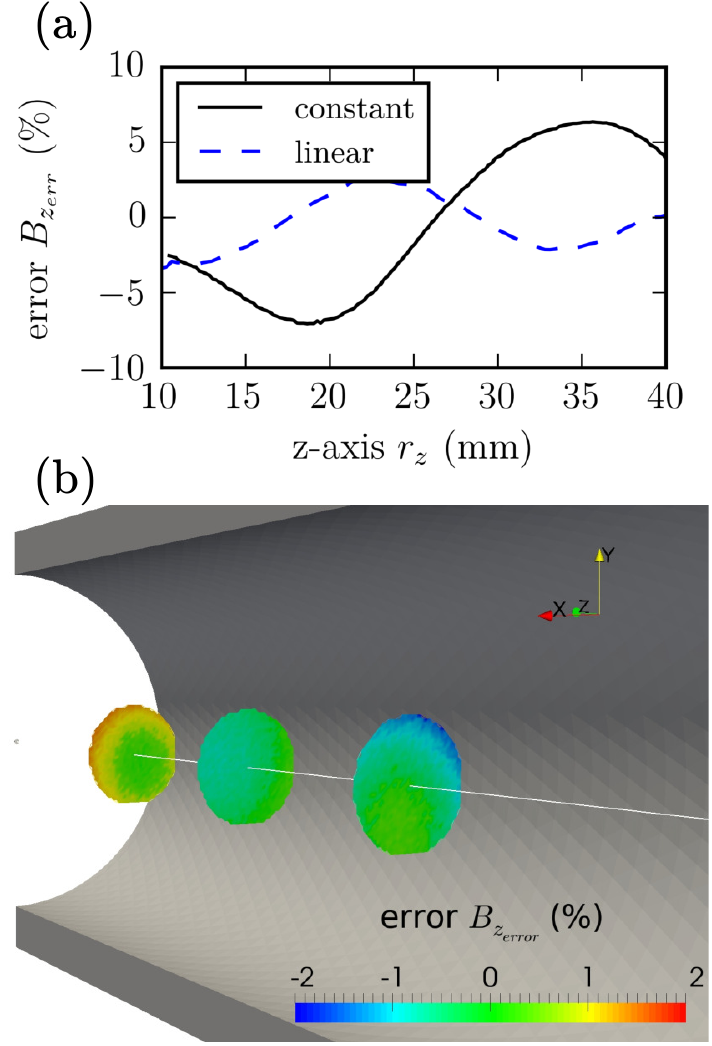}
	\caption{Errors of the printed magnets for a predefined stray field. (a) Error between the measured stray field and the inverse stray field simulation along the $z$-axis $r_z$ for the linear and constant stray field generator magnets. (b) Deviation of the eccentric stray field $B_z$ within a radius of $r=2.5$~mm on three planes ($r_z=15,~25,~35$\,mm).}
	\label{fig:density_print_error}
\end{figure}
The error between measurements and simulations are plotted in Fig.~\ref{fig:density_print_error}(a). The error changes along the $z$-axis and it is around 6\,\% for the constant and 4\,\% for the linear design. An other important feature of this magnetic design is the independence of eccentric measurements along the $z$-axis. Fig.~\ref{fig:density_print_error}(b) shows a plot of the error of eccentric measurements ($r$=2.5\,mm) of $B_z$ at three different planes ($r_z=15,~25,~35$\,mm) inside the hollow cylinder. The error is lower than 2\,\%. A picture of one of the printed magnet is presented in Fig.~\ref{fig:density_print}(d).

\section{Discussion}
Additive manufacturing of polymer bonded magnets has the advantage to manufacture magnets with a minimum of cost and time. This article presents a method to 3D print permanent magnets with a variable compound density distribution along the printing direction. With a commercially available end-user 3D printer and a mixing extruder, polymer bonded magnetic filament can be mixed with a pure PA12 filament. Hysteresis measurements with different filler fractions are performed to get a relation of remanence and compound density. The remanence decreases linearly with compound density.

To deduct the quality of the prints, an inverse stray field simulation framework is developed. No unique solution exists for this kind of inverse problem. Therefore, Tikhonov regularization is used to find reasonable results for the optimization problem. A cuboid with an absolute value function of the magnetic density is printed and the inverse stray field code is benchmarked with some measurements.

The inverse stray field code can be used to simulate magnets with a predefined target stray field in a given region outside of the magnet. The optimal magnetization density of a hollow cylinder for a targeted constant or linear stray field is computed. The magnetic compound density distribution along the $z$-axis is optimized and printed with our setup. Detailed stray field measurements shows an excellent agreement between simulation and measurement.  Eccentric stray field measurements shows a low dependence of the sensor position. This is an important aspect for linear position measurements.

With this setup and simulation framework the manufacturing of magnets are possible which are impossible to create with conventional methods. It can be used to create magnets with a specific stray field distribution for various applications. The simulation method can also be also used to improve the performance of multipolar polymer bonded magnets by injection molding. 

At the moment, only prints with a variable magnetic compound density along the $z$-axis are possible. This restriction should be rescinded by an improved slicing program. A big influence on the quality of the printed structures has the filament diameter, because with a constant feeding rate the volume flow trough the nozzle varied which leads to a patchy printing result. Here is a potential for improvement to reduce the error between simulation and measurement. Also studies of how the volumetric mass density of the printed magnets can be improved are subject to further research.

\section{Methods}
\subsection{Printing/Simulation Models}
The models for the 3D printing process and the simulation results are created in Salome 7.6. Meshing of the simulation models are performed in Salome 7.6 with the Netgen algorithm and tetrahedron elements (Supplementary Fig.~3) \cite{netgen}. Converting the .med Salome output file to the FEniCS .xml format is performed with Gmsh 2.10.1. For the manufacturing of the objects the STL data from Salome are sliced using the Slic3r software. The resulting G-code was further modified by a customized Python script to print objects with a layer depended magnetic compound density distribution.

\subsection{Stray Field Simulation}

 In a simply connected domain without current the stray field $\vec{h_m}$ of a magnetic body is \cite{jackson}
 \begin{equation}
  \vec{h_m}=-\nabla u
 \end{equation}
 with the magnetic scalar potential $u$:
  \begin{equation}
   \Delta u = \nabla \vec{M} \,\,\, \mathrm{in} \, \Omega
  \end{equation}
 and the Dirichlet boundary condition
 \begin{equation}
  u = 0  \,\,\, \mathrm{on} \, \partial \Omega
 \end{equation}

where $\vec{M}$ is the magnetization, and $\Omega=\Omega_m \cup \Omega_h \cup \Omega_a$ the different regions (Supplementary Fig. 3).

This PDE is solved by using a FEM implementation based on FEniCS2016.1 \cite{fenics}. The outer air region $\Omega_a$ is necessary to approximate the far field of the potential $u$ in $\Omega$. Air regions $\Omega_a$ which is around five-times larger than the magnetic region $\Omega_m$, gives a good compromise between accuracy of the stray field calculation and computing time. To reduce computing time the mesh size is coarser at the edge of the outer air region.

\subsection{Printer}
For the printing process the conventional end-user 3D printer Builder from Code P is chosen. This printer works by use of the fusing deposition modeling (FDM) principle. This system creates the object layer by layer by a meltable thermoplastic. It has a maximum building size of 220$\times$210$\times$164\,mm$^3$ (L$\times$W$\times$H). Structures with a layer height resolution between 0.05 and 0.3\,mm can be printed.  The printing speed ranges from 10 to 80\,mm/s, and the traveling speed from 10 to 200\,mm/s. The nozzle diameter is 0.4\,mm, and by the means of a dual feed extruder two different compound materials can be mixed, or defined region of the object can be printed with different materials. The maximum nozzle temperature is 260\,$^\circ$C. For a better adhesion of the printed objects the printer bed can be heated up to 80\,$^\circ$C. The optimal printing parameter for our setup and our magnetic compound filaments are listed in Tab. \ref{tab:print_parameter}.
\begin{table}[htb]
\centering
\begin{tabular}{m{2.8cm}|m{4cm}}
\textbf{Parameter} & \textbf{Value}\\\hline
Extruder temp. & 260\,$^\circ$C \\
Layer height & 0.15\,mm \\
Printer speed & 20\,mm/s \\
Fill density & 100\,\% \\
Fill pattern & Rectilinear \\
Bed adhesion & Kapton tape with a layer of Polyvinyl acetate (PVA) \\
Bed temp. & 60\,$^\circ$C \\
\end{tabular}
\caption{Best empirically found printer parameter for the magnetic compound material.}
\label{tab:print_parameter}
\end{table}

\subsection{Filament Manufacturing}
The polymer bonded magnetic compound consists of polyamide 12 (PA12 or also called as Nylon 12) from Polyking (221-TR) and magnetically isotropic powder MQP-S-11-9 with the chemical composition NdPrFeCoTiZrB from Magnequench Corporation. These source materials are compounded and extruded into suitable filaments in the desired ratio of 85\,wt.-\% MQP-S-11-9 powder and 15\,wt.-\% PA12. The extrusion is performed at University of Leoben with a Leistritz ZSE 18 HPe-48D twin-screw extruder. The materials are dried at 80\,$^\circ$C for 8 hours. The four heating zones of the twin-screw are temperate. The feed section is the coolest with 80\,$^\circ$C and the temperature increases up to the shaping die which has a temperature of 260\,$^\circ$C. The round orifice of the die has a diameter of 1.75\,mm. The hot extrudate is hauled off and cooled by a cooled conveyor belt.. The diameter and tolerances of the filament is controlled by a Sikora Laser Series 2000 diameter measuring system. The extrusion speed is adjusted to get a filament with a diameter of 1.75\,mm. The manufactured filament is spooled with a diameter of around 0.5\,m to avoid breaking of the brittle magnetic filaments.

\subsection{Material Characterization}
The fraction of NdFeB particles in the PA12 matrix is measured by thermogravimetric analysis (TGA). The model TGA~2050 from TA-Instruments has a resolution of 0.2\,\textmu g and a temperature range of 25-1000\,$^\circ$C. In our case a heating rate of 10\,K/min and a nitrogen atmosphere to avoid oxidation of the particles is used. TGA measurement yields a filler content of 85\,wt.\%. Hysteresis measurements are performed for different magnetic compound fractions $\varrho_m$. With the dual extruder of our printer cubes with a size of 7\,mm are printed and afterwards post processed to obtain cubes with a length of $a$ of 5$\pm$0.02\,mm. The hysteresis is measured by Pulsed Field Magnetometry (PFM) (Hirst PFM11) \cite{pfm2, pfm}. All measurements are carried out with the same parameters - temperature of 297\,K and a magnetic field up to 4\,T peak field. The internal field is $H_{\text{int}}=H_{\text{ext}}-N/\mu_0 J$. Where $H_{\text{ext}}$ is the external field, $N$ is the average demagnetisation factor for a cube ($N=1/3$) \cite{demag}, and $J$ is the material polarization. The morphology of the NdFeB particles is identified by Scanning Electron Microscope (FEI Quanta 200 FEG) images. The samples are Au coated with a Sputter Coater Quorum (Q150T S). The particles are of spherical shape with a diameter of approximately 45$\pm$20\,\textmu m (Supplementary Fig.1).

\subsection{Magnetization}
The objects with a variable magnetic compound density are magnetized inside an electromagnet. It is a self-build water cooled electromagnet, and it is powered by a low voltage power supply (Siemens NTN 35000-200). Maximum output current is 150\,A with a operating voltage of 200\,V. This setup has a maximum magnetic flux density inside the electromagnet of 1.9\,T in in a permanent operation mode. The gap between the pole shoes is 50\,mm.

\subsection{Stray Field Measurement}
To measure the stray field of the printed permanent magnet, the 3D printer is upgraded to a full 3D magnetic flux density measurement system. As sensing device, a 3D Hall sensor TLV493D-A186 from Infineon is used.  A Genuino 101 microcontroller is programmed to read out the components of $\vec{B}$ with a frequency of 3\,kHz. The sensor has a measurement range of $\pm$130\,mT, and a measured detectivity of 40\,\textmu T/$\sqrt{\mathrm{Hz}}$ for static magnetic fields. A Python script controls the movement of the 3D printer and saves the stray field measurement data for the actual position of the sensor. This setup has a spatial resolution of 0.05\,mm along the $z$-axis and 0.1\,mm along the $x$ and $y$-axis. To skip an elaborate adjusting and alignment of the sensor a calibration method on detailed stray field simulation is used \cite{pub_16_1_apl}. With this method the angles, sensitivity, and the offset of the sensor can be calibrated. In our case the sensor is simply attached to the extruder head with a self printed suspension without any adjustment (Supplementary Fig.~5). With this setup the stray field  can be scanned in 1D, 2D, and 3D around a complex magnetic structure.

\bibliographystyle{aipnum4-1}
%\bibliography{pub16_2}

\begin{thebibliography}{24}%
\makeatletter
\providecommand \@ifxundefined [1]{%
 \@ifx{#1\undefined}
}%
\providecommand \@ifnum [1]{%
 \ifnum #1\expandafter \@firstoftwo
 \else \expandafter \@secondoftwo
 \fi
}%
\providecommand \@ifx [1]{%
 \ifx #1\expandafter \@firstoftwo
 \else \expandafter \@secondoftwo
 \fi
}%
\providecommand \natexlab [1]{#1}%
\providecommand \enquote  [1]{``#1''}%
\providecommand \bibnamefont  [1]{#1}%
\providecommand \bibfnamefont [1]{#1}%
\providecommand \citenamefont [1]{#1}%
\providecommand \href@noop [0]{\@secondoftwo}%
\providecommand \href [0]{\begingroup \@sanitize@url \@href}%
\providecommand \@href[1]{\@@startlink{#1}\@@href}%
\providecommand \@@href[1]{\endgroup#1\@@endlink}%
\providecommand \@sanitize@url [0]{\catcode `\\12\catcode `\$12\catcode
  `\&12\catcode `\#12\catcode `\^12\catcode `\_12\catcode `\%12\relax}%
\providecommand \@@startlink[1]{}%
\providecommand \@@endlink[0]{}%
\providecommand \url  [0]{\begingroup\@sanitize@url \@url }%
\providecommand \@url [1]{\endgroup\@href {#1}{\urlprefix }}%
\providecommand \urlprefix  [0]{URL }%
\providecommand \Eprint [0]{\href }%
\providecommand \doibase [0]{http://dx.doi.org/}%
\providecommand \selectlanguage [0]{\@gobble}%
\providecommand \bibinfo  [0]{\@secondoftwo}%
\providecommand \bibfield  [0]{\@secondoftwo}%
\providecommand \translation [1]{[#1]}%
\providecommand \BibitemOpen [0]{}%
\providecommand \bibitemStop [0]{}%
\providecommand \bibitemNoStop [0]{.\EOS\space}%
\providecommand \EOS [0]{\spacefactor3000\relax}%
\providecommand \BibitemShut  [1]{\csname bibitem#1\endcsname}%
\let\auto@bib@innerbib\@empty
%</preamble>
\bibitem [{\citenamefont {Guo}\ and\ \citenamefont {Leu}(2013)}]{3d-print}%
  \BibitemOpen
  \bibfield  {author} {\bibinfo {author} {\bibfnamefont {N.}~\bibnamefont
  {Guo}}\ and\ \bibinfo {author} {\bibfnamefont {M.~C.}\ \bibnamefont {Leu}},\
  }\href@noop {} {\bibfield  {journal} {\bibinfo  {journal} {Frontiers of
  Mechanical Engineering}\ }\textbf {\bibinfo {volume} {8}},\ \bibinfo {pages}
  {215} (\bibinfo {year} {2013})}\BibitemShut {NoStop}%
\bibitem [{\citenamefont {Wohlers}, \citenamefont {Caffrey},\ and\
  \citenamefont {Campbell}(2016)}]{wohlers}%
  \BibitemOpen
  \bibfield  {author} {\bibinfo {author} {\bibfnamefont {T.~T.}\ \bibnamefont
  {Wohlers}}, \bibinfo {author} {\bibfnamefont {T.}~\bibnamefont {Caffrey}}, \
  and\ \bibinfo {author} {\bibfnamefont {R.~I.}\ \bibnamefont {Campbell}},\
  }\href@noop {} {\emph {\bibinfo {title} {Wohlers Report 2016: 3D Printing and
  Additive Manufacturing State of the Industry Annual Worldwide Progress
  Report}}}\ (\bibinfo  {publisher} {Wohlers Associates},\ \bibinfo {year}
  {2016})\BibitemShut {NoStop}%
\bibitem [{\citenamefont {Ma}\ \emph {et~al.}(2002)\citenamefont {Ma},
  \citenamefont {Herchenroeder}, \citenamefont {Smith}, \citenamefont {Suda},
  \citenamefont {Brown},\ and\ \citenamefont {Chen}}]{recent_devel}%
  \BibitemOpen
  \bibfield  {author} {\bibinfo {author} {\bibfnamefont {B.}~\bibnamefont
  {Ma}}, \bibinfo {author} {\bibfnamefont {J.}~\bibnamefont {Herchenroeder}},
  \bibinfo {author} {\bibfnamefont {B.}~\bibnamefont {Smith}}, \bibinfo
  {author} {\bibfnamefont {M.}~\bibnamefont {Suda}}, \bibinfo {author}
  {\bibfnamefont {D.}~\bibnamefont {Brown}}, \ and\ \bibinfo {author}
  {\bibfnamefont {Z.}~\bibnamefont {Chen}},\ }\href@noop {} {\bibfield
  {journal} {\bibinfo  {journal} {Journal of Magnetism and Magnetic Materials}\
  }\textbf {\bibinfo {volume} {239}},\ \bibinfo {pages} {418 } (\bibinfo {year}
  {2002})}\BibitemShut {NoStop}%
\bibitem [{\citenamefont {Ormerod}\ and\ \citenamefont
  {Constantinides}(1997)}]{bonded_mag_future}%
  \BibitemOpen
  \bibfield  {author} {\bibinfo {author} {\bibfnamefont {J.}~\bibnamefont
  {Ormerod}}\ and\ \bibinfo {author} {\bibfnamefont {S.}~\bibnamefont
  {Constantinides}},\ }\href@noop {} {\bibfield  {journal} {\bibinfo  {journal}
  {Journal of Applied Physics}\ }\textbf {\bibinfo {volume} {81}},\ \bibinfo
  {pages} {4816} (\bibinfo {year} {1997})}\BibitemShut {NoStop}%
\bibitem [{\citenamefont {Elian}\ and\ \citenamefont {Theuss}(2014)}]{ibb}%
  \BibitemOpen
  \bibfield  {author} {\bibinfo {author} {\bibfnamefont {K.}~\bibnamefont
  {Elian}}\ and\ \bibinfo {author} {\bibfnamefont {H.}~\bibnamefont {Theuss}},\
  }in\ \href@noop {} {\emph {\bibinfo {booktitle} {Electronics
  System-Integration Technology Conference (ESTC), 2014}}}\ (\bibinfo {year}
  {2014})\ pp.\ \bibinfo {pages} {1--5}\BibitemShut {NoStop}%
\bibitem [{\citenamefont {Eimeke}\ \emph {et~al.}(2008)\citenamefont {Eimeke},
  \citenamefont {Gardocki}, \citenamefont {Ehrenstein},\ and\ \citenamefont
  {Drummer}}]{diss_drummer}%
  \BibitemOpen
  \bibfield  {author} {\bibinfo {author} {\bibfnamefont {S.}~\bibnamefont
  {Eimeke}}, \bibinfo {author} {\bibfnamefont {A.}~\bibnamefont {Gardocki}},
  \bibinfo {author} {\bibfnamefont {G.}~\bibnamefont {Ehrenstein}}, \ and\
  \bibinfo {author} {\bibfnamefont {D.}~\bibnamefont {Drummer}},\ }\href@noop
  {} {\bibfield  {journal} {\bibinfo  {journal} {Journal of Plastics
  Technology}\ }\textbf {\bibinfo {volume} {5}} (\bibinfo {year}
  {2008})}\BibitemShut {NoStop}%
\bibitem [{\citenamefont {Gonzalez-Gutierrez}\ \emph
  {et~al.}(2016)\citenamefont {Gonzalez-Gutierrez}, \citenamefont {Duretek},
  \citenamefont {Kukla}, \citenamefont {Polj{\v{s}}ak}, \citenamefont {Bek},
  \citenamefont {Emri},\ and\ \citenamefont {Holzer}}]{models_visc}%
  \BibitemOpen
  \bibfield  {author} {\bibinfo {author} {\bibfnamefont {J.}~\bibnamefont
  {Gonzalez-Gutierrez}}, \bibinfo {author} {\bibfnamefont {I.}~\bibnamefont
  {Duretek}}, \bibinfo {author} {\bibfnamefont {C.}~\bibnamefont {Kukla}},
  \bibinfo {author} {\bibfnamefont {A.}~\bibnamefont {Polj{\v{s}}ak}}, \bibinfo
  {author} {\bibfnamefont {M.}~\bibnamefont {Bek}}, \bibinfo {author}
  {\bibfnamefont {I.}~\bibnamefont {Emri}}, \ and\ \bibinfo {author}
  {\bibfnamefont {C.}~\bibnamefont {Holzer}},\ }\href@noop {} {\bibfield
  {journal} {\bibinfo  {journal} {Metals}\ }\textbf {\bibinfo {volume} {6}},\
  \bibinfo {pages} {129} (\bibinfo {year} {2016})}\BibitemShut {NoStop}%
\bibitem [{\citenamefont {Huber}\ \emph {et~al.}(2016)\citenamefont {Huber},
  \citenamefont {Abert}, \citenamefont {Bruckner}, \citenamefont {Groenefeld},
  \citenamefont {Muthsam}, \citenamefont {Schuschnigg}, \citenamefont {Sirak},
  \citenamefont {Thanhoffer}, \citenamefont {Teliban}, \citenamefont {Vogler},
  \citenamefont {Windl},\ and\ \citenamefont {Suess}}]{pub_16_1_apl}%
  \BibitemOpen
  \bibfield  {author} {\bibinfo {author} {\bibfnamefont {C.}~\bibnamefont
  {Huber}}, \bibinfo {author} {\bibfnamefont {C.}~\bibnamefont {Abert}},
  \bibinfo {author} {\bibfnamefont {F.}~\bibnamefont {Bruckner}}, \bibinfo
  {author} {\bibfnamefont {M.}~\bibnamefont {Groenefeld}}, \bibinfo {author}
  {\bibfnamefont {O.}~\bibnamefont {Muthsam}}, \bibinfo {author} {\bibfnamefont
  {S.}~\bibnamefont {Schuschnigg}}, \bibinfo {author} {\bibfnamefont
  {K.}~\bibnamefont {Sirak}}, \bibinfo {author} {\bibfnamefont
  {R.}~\bibnamefont {Thanhoffer}}, \bibinfo {author} {\bibfnamefont
  {I.}~\bibnamefont {Teliban}}, \bibinfo {author} {\bibfnamefont
  {C.}~\bibnamefont {Vogler}}, \bibinfo {author} {\bibfnamefont
  {R.}~\bibnamefont {Windl}}, \ and\ \bibinfo {author} {\bibfnamefont
  {D.}~\bibnamefont {Suess}},\ }\href@noop {} {\bibfield  {journal} {\bibinfo
  {journal} {Applied Physics Letters}\ }\textbf {\bibinfo {volume} {109}}
  (\bibinfo {year} {2016})}\BibitemShut {NoStop}%
\bibitem [{\citenamefont {Li}\ \emph {et~al.}(2016)\citenamefont {Li},
  \citenamefont {Tirado}, \citenamefont {Nlebedim}, \citenamefont {Rios},
  \citenamefont {Post}, \citenamefont {Kunc}, \citenamefont {Lowden},
  \citenamefont {Lara-Curzio}, \citenamefont {Fredette}, \citenamefont
  {Ormerod}, \citenamefont {Lograsso},\ and\ \citenamefont
  {Paranthaman}}]{baam}%
  \BibitemOpen
  \bibfield  {author} {\bibinfo {author} {\bibfnamefont {L.}~\bibnamefont
  {Li}}, \bibinfo {author} {\bibfnamefont {A.}~\bibnamefont {Tirado}}, \bibinfo
  {author} {\bibfnamefont {I.~C.}\ \bibnamefont {Nlebedim}}, \bibinfo {author}
  {\bibfnamefont {O.}~\bibnamefont {Rios}}, \bibinfo {author} {\bibfnamefont
  {B.}~\bibnamefont {Post}}, \bibinfo {author} {\bibfnamefont {V.}~\bibnamefont
  {Kunc}}, \bibinfo {author} {\bibfnamefont {R.~R.}\ \bibnamefont {Lowden}},
  \bibinfo {author} {\bibfnamefont {E.}~\bibnamefont {Lara-Curzio}}, \bibinfo
  {author} {\bibfnamefont {R.}~\bibnamefont {Fredette}}, \bibinfo {author}
  {\bibfnamefont {J.}~\bibnamefont {Ormerod}}, \bibinfo {author} {\bibfnamefont
  {T.~A.}\ \bibnamefont {Lograsso}}, \ and\ \bibinfo {author} {\bibfnamefont
  {M.~P.}\ \bibnamefont {Paranthaman}},\ }\href@noop {} {\bibfield  {journal}
  {\bibinfo  {journal} {Scientific Reports}\ }\textbf {\bibinfo {volume} {6}},\
  \bibinfo {pages} {36212} (\bibinfo {year} {2016})}\BibitemShut {NoStop}%
\bibitem [{\citenamefont {Burkhardt}(2015)}]{reduction_ndfeb}%
  \BibitemOpen
  \bibfield  {author} {\bibinfo {author} {\bibfnamefont {C.}~\bibnamefont
  {Burkhardt}},\ }in\ \href@noop {} {\emph {\bibinfo {booktitle} {Pforzheimer
  Werkstofftag}}}\ (\bibinfo {year} {2015})\BibitemShut {NoStop}%
\bibitem [{\citenamefont {Abert}\ \emph {et~al.}(2013)\citenamefont {Abert},
  \citenamefont {Exl}, \citenamefont {Selke}, \citenamefont {Drews},\ and\
  \citenamefont {Schrefl}}]{stray_field_fem}%
  \BibitemOpen
  \bibfield  {author} {\bibinfo {author} {\bibfnamefont {C.}~\bibnamefont
  {Abert}}, \bibinfo {author} {\bibfnamefont {L.}~\bibnamefont {Exl}}, \bibinfo
  {author} {\bibfnamefont {G.}~\bibnamefont {Selke}}, \bibinfo {author}
  {\bibfnamefont {A.}~\bibnamefont {Drews}}, \ and\ \bibinfo {author}
  {\bibfnamefont {T.}~\bibnamefont {Schrefl}},\ }\href@noop {} {\bibfield
  {journal} {\bibinfo  {journal} {Journal of Magnetism and Magnetic Materials}\
  }\textbf {\bibinfo {volume} {326}},\ \bibinfo {pages} {176 } (\bibinfo {year}
  {2013})}\BibitemShut {NoStop}%
\bibitem [{\citenamefont {Fredkin}\ and\ \citenamefont
  {Koehler}(1990)}]{fredkin}%
  \BibitemOpen
  \bibfield  {author} {\bibinfo {author} {\bibfnamefont {D.~R.}\ \bibnamefont
  {Fredkin}}\ and\ \bibinfo {author} {\bibfnamefont {T.~R.}\ \bibnamefont
  {Koehler}},\ }\href@noop {} {\bibfield  {journal} {\bibinfo  {journal} {IEEE
  Transactions on Magnetics}\ }\textbf {\bibinfo {volume} {26}},\ \bibinfo
  {pages} {415} (\bibinfo {year} {1990})}\BibitemShut {NoStop}%
\bibitem [{\citenamefont {{Bruckner}}\ \emph {et~al.}(2016)\citenamefont
  {{Bruckner}}, \citenamefont {{Abert}}, \citenamefont {{Wautischer}},
  \citenamefont {{Huber}}, \citenamefont {{Vogler}}, \citenamefont {{Hinze}},\
  and\ \citenamefont {{Suess}}}]{inverse_flo}%
  \BibitemOpen
  \bibfield  {author} {\bibinfo {author} {\bibfnamefont {F.}~\bibnamefont
  {{Bruckner}}}, \bibinfo {author} {\bibfnamefont {C.}~\bibnamefont {{Abert}}},
  \bibinfo {author} {\bibfnamefont {G.}~\bibnamefont {{Wautischer}}}, \bibinfo
  {author} {\bibfnamefont {C.}~\bibnamefont {{Huber}}}, \bibinfo {author}
  {\bibfnamefont {C.}~\bibnamefont {{Vogler}}}, \bibinfo {author}
  {\bibfnamefont {M.}~\bibnamefont {{Hinze}}}, \ and\ \bibinfo {author}
  {\bibfnamefont {D.}~\bibnamefont {{Suess}}},\ }\href@noop {} {\bibfield
  {journal} {\bibinfo  {journal} {ArXiv e-prints}\ } (\bibinfo {year}
  {2016})},\ \Eprint {http://arxiv.org/abs/1609.00060} {arXiv:1609.00060
  [physics.comp-ph]} \BibitemShut {NoStop}%
\bibitem [{\citenamefont {Logg}(2013)}]{fenics}%
  \BibitemOpen
  \bibfield  {author} {\bibinfo {author} {\bibfnamefont {A.}~\bibnamefont
  {Logg}},\ }\href@noop {} {\emph {\bibinfo {title} {Automated Solution of
  Differential Equations by the Finite Element Method (Lecture Notes in
  Computational Science and Engineering)}}}\ (\bibinfo  {publisher}
  {Springer},\ \bibinfo {year} {2013})\BibitemShut {NoStop}%
\bibitem [{\citenamefont {{Funke}}\ and\ \citenamefont
  {{Farrell}}(2013)}]{dolfin}%
  \BibitemOpen
  \bibfield  {author} {\bibinfo {author} {\bibfnamefont {S.~W.}\ \bibnamefont
  {{Funke}}}\ and\ \bibinfo {author} {\bibfnamefont {P.~E.}\ \bibnamefont
  {{Farrell}}},\ }\href@noop {} {\bibfield  {journal} {\bibinfo  {journal}
  {ArXiv e-prints}\ } (\bibinfo {year} {2013})},\ \Eprint
  {http://arxiv.org/abs/1302.3894} {arXiv:1302.3894} \BibitemShut {NoStop}%
\bibitem [{\citenamefont {Tikhonov}\ \emph {et~al.}(1995)\citenamefont
  {Tikhonov}, \citenamefont {Goncharsky}, \citenamefont {Stepanov},\ and\
  \citenamefont {Yagola}}]{inverse}%
  \BibitemOpen
  \bibfield  {author} {\bibinfo {author} {\bibfnamefont {A.}~\bibnamefont
  {Tikhonov}}, \bibinfo {author} {\bibfnamefont {A.}~\bibnamefont
  {Goncharsky}}, \bibinfo {author} {\bibfnamefont {V.}~\bibnamefont
  {Stepanov}}, \ and\ \bibinfo {author} {\bibfnamefont {A.~G.}\ \bibnamefont
  {Yagola}},\ }\href@noop {} {\emph {\bibinfo {title} {Numerical Methods for
  the Solution of Ill-Posed Problems (Mathematics and Its Applications)}}}\
  (\bibinfo  {publisher} {Springer},\ \bibinfo {year} {1995})\BibitemShut
  {NoStop}%
\bibitem [{\citenamefont {O'Leary}\ and\ \citenamefont
  {Prost}(1993)}]{l-curve}%
  \BibitemOpen
  \bibfield  {author} {\bibinfo {author} {\bibnamefont {O'Leary}}\ and\
  \bibinfo {author} {\bibfnamefont {D.}~\bibnamefont {Prost}},\ }\href@noop {}
  {\bibfield  {journal} {\bibinfo  {journal} {Society for Industrial and
  Applied Mathematics}\ }\textbf {\bibinfo {volume} {14}},\ \bibinfo {pages}
  {1287} (\bibinfo {year} {1993})}\BibitemShut {NoStop}%
\bibitem [{\citenamefont {W{\"a}chter}\ and\ \citenamefont
  {Biegler}(2006)}]{ipopt}%
  \BibitemOpen
  \bibfield  {author} {\bibinfo {author} {\bibfnamefont {A.}~\bibnamefont
  {W{\"a}chter}}\ and\ \bibinfo {author} {\bibfnamefont {L.~T.}\ \bibnamefont
  {Biegler}},\ }\href@noop {} {\bibfield  {journal} {\bibinfo  {journal}
  {Mathematical Programming}\ }\textbf {\bibinfo {volume} {106}},\ \bibinfo
  {pages} {25} (\bibinfo {year} {2006})}\BibitemShut {NoStop}%
\bibitem [{\citenamefont {Ortner}(2015)}]{linear}%
  \BibitemOpen
  \bibfield  {author} {\bibinfo {author} {\bibfnamefont {M.}~\bibnamefont
  {Ortner}},\ }in\ \href@noop {} {\emph {\bibinfo {booktitle} {2015 9th
  International Conference on Sensing Technology (ICST)}}}\ (\bibinfo {year}
  {2015})\ pp.\ \bibinfo {pages} {359--364}\BibitemShut {NoStop}%
\bibitem [{\citenamefont {Sch{\"o}berl}(1997)}]{netgen}%
  \BibitemOpen
  \bibfield  {author} {\bibinfo {author} {\bibfnamefont {J.}~\bibnamefont
  {Sch{\"o}berl}},\ }\href@noop {} {\bibfield  {journal} {\bibinfo  {journal}
  {Computing and Visualization in Science}\ }\textbf {\bibinfo {volume} {1}},\
  \bibinfo {pages} {41} (\bibinfo {year} {1997})}\BibitemShut {NoStop}%
\bibitem [{\citenamefont {Jackson}(1999)}]{jackson}%
  \BibitemOpen
  \bibfield  {author} {\bibinfo {author} {\bibfnamefont {J.}~\bibnamefont
  {Jackson}},\ }\href@noop {} {\emph {\bibinfo {title} {Classical
  electrodynamics}}}\ (\bibinfo  {publisher} {Wiley},\ \bibinfo {address} {New
  York},\ \bibinfo {year} {1999})\BibitemShut {NoStop}%
\bibitem [{\citenamefont {Groesinger}(2008)}]{pfm2}%
  \BibitemOpen
  \bibfield  {author} {\bibinfo {author} {\bibfnamefont {R.}~\bibnamefont
  {Groesinger}},\ }\href@noop {} {\bibfield  {journal} {\bibinfo  {journal}
  {Journal of Electrical Engineering}\ }\textbf {\bibinfo {volume} {59}},\
  \bibinfo {pages} {15} (\bibinfo {year} {2008})}\BibitemShut {NoStop}%
\bibitem [{\citenamefont {Fiorillo}\ \emph {et~al.}(2007)\citenamefont
  {Fiorillo}, \citenamefont {Beatrice}, \citenamefont {Bottauscio},\ and\
  \citenamefont {Patroi}}]{pfm}%
  \BibitemOpen
  \bibfield  {author} {\bibinfo {author} {\bibfnamefont {F.}~\bibnamefont
  {Fiorillo}}, \bibinfo {author} {\bibfnamefont {C.}~\bibnamefont {Beatrice}},
  \bibinfo {author} {\bibfnamefont {O.}~\bibnamefont {Bottauscio}}, \ and\
  \bibinfo {author} {\bibfnamefont {E.}~\bibnamefont {Patroi}},\ }\href@noop {}
  {\bibfield  {journal} {\bibinfo  {journal} {IEEE Transactions on Magnetics}\
  }\textbf {\bibinfo {volume} {43}},\ \bibinfo {pages} {3159} (\bibinfo {year}
  {2007})}\BibitemShut {NoStop}%
\bibitem [{\citenamefont {Aharoni}(1998)}]{demag}%
  \BibitemOpen
  \bibfield  {author} {\bibinfo {author} {\bibfnamefont {A.}~\bibnamefont
  {Aharoni}},\ }\href@noop {} {\bibfield  {journal} {\bibinfo  {journal}
  {Journal of Applied Physics}\ }\textbf {\bibinfo {volume} {83}},\ \bibinfo
  {pages} {3432} (\bibinfo {year} {1998})}\BibitemShut {NoStop}%
\end{thebibliography}


\section{Acknowledgement}
The support from CD-Laboratory AMSEN (financed by the Austrian Federal Ministry of Economy, Family and Youth, the National Foundation for Research, Technology and Development) is acknowledged. The authors would like to thank Magnetfabrik Bonn GmbH for the provision of the compound material, and to Montanuniversitaet Leoben for the extrusion of the filaments. The SEM and sample preparations are carried out using facilities at the University Service Centre for Transmission Electron Microscopy, TU-Vienna, Austria. The computational results presented have been achieved using the Vienna Scientific Cluster (VSC).

%\section{Author Contributions}
%C.H., M.G., I.T., and D.S conceived the idea of 3D printing permanent magnets. C.H. and D.S. designed the print examples. C.H. and S.S. compounded and extruded the filaments. C.H. and R.W. designed the stray field measurement system, and carried it out. C.H. characterized the material. C.H., C.A., F.B., C.V., and G.W. programmed the inverse stray field framework with Fenics and dolfin-adjoint. The manuscript was written by C.H. All authors reviewed the manuscript.

\end{document}